\documentclass[lettersize,journal]{IEEEtran}
\usepackage{amsmath,amssymb,amsfonts}
\usepackage{graphicx}
\usepackage{subcaption}
\usepackage{xcolor}
\usepackage{tabularx}
\usepackage{xspace}
\usepackage[inline]{enumitem} 
\usepackage{booktabs}
\usepackage{multirow}
\usepackage{hyperref}
\graphicspath{{}}
\usepackage{cite}
\usepackage{color, colortbl}
\usepackage{caption}
\usepackage[many]{tcolorbox}
\usepackage{textcomp}
\usepackage{tabularx}
\usepackage{subcaption}
\usepackage{ragged2e}
\usepackage{blindtext}
\usepackage{rotating}

\definecolor{main}{HTML}{5989cf}    
\definecolor{sub}{HTML}{cde4ff}     

\tcbset{
    sharp corners,
    colback = white,
    before skip = 0.2cm,    
    after skip = 0.5cm      
}                           

\newcommand{\hypohyper}[0]{hyponym-hypernym\xspace}

\newcommand{\relation}[2]{#1$\rightarrow$#2\xspace}

\usepackage{cite}

\newcommand{\lbl}[1]{`\textit{#1}'\xspace}
\newcommand*{\ie}{i.e.,\@\xspace}
\newcommand*{\eg}{e.g.,\@\xspace}

\newcommand*{\etal}{et al.\@\xspace}

\newcommand*\rot{\rotatebox{90}}
\newcommand*\OK{\ding{51}}

\usepackage{pifont}

\newtcolorbox[auto counter]{llmprompt}[1][]{%
    title=LLM Prompt~\thetcbcounter, arc=1pt,outer arc=1pt,float,fontupper=\small, #1}
	
\newtcolorbox{boxH}{
    colback = sub, 
    colframe = main, 
    boxrule = 0pt, 
    leftrule = 6pt 
}

\ifCLASSINFOpdf
\else
\fi
\usepackage{tikz}
\usetikzlibrary{shapes.geometric, arrows}

\tikzstyle{startstop} = [rectangle, rounded corners, minimum width=3cm, minimum height=1cm,text centered, draw=black, fill=red!30]
\tikzstyle{process} = [rectangle, minimum width=3cm, minimum height=1cm, text centered, draw=black, fill=orange!30]
\tikzstyle{decision} = [diamond, minimum width=3cm, minimum height=1cm, text centered, draw=black, fill=green!30]
\tikzstyle{arrow} = [thick,->,>=stealth]

\begin{document}
%
\title{Automatic Bottom-Up Taxonomy Construction: \\ A Software Application Domain Study}
%
%
%

\author{Cezar~Sas 
        and~Andrea~Capiluppi
\thanks{}
\thanks{Manuscript submitted September 18, 2024.}
}

%
%

\markboth{Journal of \LaTeX\ Class Files,~Vol.~14, No.~8, August~2015}%
{Shell \MakeLowercase{\textit{et al.}}: Bare Demo of IEEEtran.cls for IEEE Journals}
%



\maketitle

\begin{abstract}

Previous research in software application domain classification has faced challenges due to the lack of a proper taxonomy that explicitly models relations between classes. As a result, current solutions are less effective for real-world usage.

This study aims to develop a comprehensive software application domain taxonomy by integrating multiple datasources and leveraging ensemble methods. The goal is to overcome the limitations of individual sources and configurations by creating a more robust, accurate, and reproducible taxonomy.

This study employs a quantitative research design involving three different datasources: an existing Computer Science Ontology (CSO), Wikidata, and large language models (LLMs). The study utilises a combination of automated and human evaluations to assess the quality of a taxonomy. The outcome measures include the number of unlinked terms, self-loops, and overall connectivity of the taxonomy. 

The results indicate that individual datasources have advantages and drawbacks: the CSO datasource showed minimal variance across different configurations, but a notable issue of missing technical terms and a high number of self-loops. The Wikipedia datasource required significant filtering during construction to improve metric performance. LLM-generated taxonomies demonstrated better performance when using context-rich prompts. An ensemble approach showed the most promise, successfully reducing the number of unlinked terms and self-loops, thus creating a more connected and comprehensive taxonomy. 

The study addresses the construction of a software application domain taxonomy relying on pre-existing resources. Our results indicate that an ensemble approach to taxonomy construction can effectively address the limitations of individual datasources. 
Future work should focus on refining the ensemble techniques and exploring additional datasources to further enhance the taxonomy's accuracy and completeness. 
\end{abstract}


%
\IEEEpeerreviewmaketitle

\section{Introduction}
%
%
%
%


\IEEEPARstart{I}n the context of software engineering, a classification is a systematic arrangement of data, software components or concepts into distinct labels based on predefined criteria. The aim of a classification is to create an organised structure that enhances understanding, management and retrieval of information, facilitating various processes such as development, debugging, maintenance and optimisation of software systems. Effective classification systems are essential for organizing and interpreting data, thus enabling informed decision-making~\cite{nickerson2013method}. This is also important for software repositories, where application domain classification~\cite{sas2023weak, Kawaguchi2006MUDABlue, sipio2020naive, izadi201repologue} can aid in software  reuse~\cite{kruger2021reuse,Biggerstaff1989reuse}, domain-specific empirical research~\cite{briand2012embracing, briand2017case}, and enhance software modularization~\cite{sarhan2022software, teymourian2022fast, yang2022enhancing}.


Existing classifications in software engineering are often built on flat structures that ignore the hierarchical relationships between terms~\cite{sas2022antipatterns}. The main issue with this approach is that it overlooks the \textit{IS-A} relationships between labels (for example, \lbl{Neural Networks} is a subdomain of \lbl{Machine Learning}). Ignoring these relationships can make a classification incomplete and potentially impair model training. This happens because different classes may have shared features (for instance, terms like 'train' or 'precision' appear in documents from both \lbl{Neural Networks} and \lbl{Machine Learning}), especially when training labels are incomplete or sparse.

In contrast, explicit hierarchies (or taxonomies) provide a classification framework that organizes and categorizes concepts, artefacts, or processes by connecting them through meaningful relationships, often visualized as hierarchical structures. Taxonomies enable the training of classifiers that balance specificity and accuracy, resulting in highly specific classifiers without sacrificing accuracy~\cite{deng2012hedging}. Additionally, examples from similar classes can be helpful for classes with few labels in a long-tailed distribution. It has been shown that classification taxonomies facilitate feature sharing and improve information transfer across different classes~\cite{wu2020solving}.

Constructing a taxonomy generally follows three common steps~\cite{hedden2016accidental}:
\begin{enumerate*}
    \item \textbf{Terms Selection}: in this step, the relevant terms are identified within the domain that will be included in the taxonomy.
    \item \textbf{Relationship Construction}: this step is focused on establishing hierarchical relationships between selected terms, organizing them into categories and subcategories to illustrate interconnections within the domain.
    \item \textbf{Evaluation}: the taxonomy must be assessed in accuracy, comprehensiveness, and usability through expert validation, empirical testing, and iterative refinement.
\end{enumerate*}

These construction steps can be carried out either manually by domain experts or automatically~\cite{wang2017survey, shang2020nettaxo, zhang2018taxogen}. The manual approach involves assembling a team of knowledgeable domain experts who collaboratively develop the taxonomy. However, finding a sufficient number of qualified experts can be difficult, and reaching a consensus among them often poses challenges, especially regarding its usability~\cite{forward2008taxonomy}. In contrast, automatic ones require high-quality domain-specific data to produce accurate and meaningful results, which might be challenging and time-consuming to obtain.

To mitigate the risks of each approach (manual and automated), and leverage the benefits of each alternative, we base the construction of our taxonomy on a hybrid solution, combining pre-existing, human-made knowledge with automatically generated knowledge. For the first step (\textit{terms selection}), we use a small initial set of terms (\eg a `seed') that leverages existing knowledge from various sources, each with unique characteristics. Using a seed allows for a bottom-up approach, making the methodology adaptable to pre-existing classifications. 

For the second step (\textit{relationship construction}), we utilize more specific resources that contain relationships between the terms, for example, related \textit{ontologies}\footnote{Our approach can work with similar or more general sets or subsets of ontologies, and does not need an exact or complete, context-aware ontology.}. Existing ontologies provide structured and curated knowledge, but they are often limited in coverage: this is usually because their authors originally created them for a smaller scope or different intent (\eg CSO~\cite{salatino2018cso} was built for academic research). Furthermore, existing taxonomies do not encompass the entire breadth of the domain, whereas our taxonomy aims to cover the whole software engineering spectrum of terms. We also expand the resulting taxonomy by using a general knowledge base to offer broader coverage and to allow more terms to be included. Lastly, we employ the generative abilities of LLMs to link the remaining unconnected terms and fill any remaining gaps.

For the final step (\textit{result evaluation}), we evaluate the resulting taxonomy using a combination of automatic and manual approaches (e.g., using human annotators). 

The contributions of this work are the following:
\begin{itemize}
\item A novel approach for constructing a bottom-up taxonomy by integrating multiple pre-existing resources, leveraging both domain-specific and general knowledge bases;
\item 
A thorough evaluation of each resource's strengths and limitations in the taxonomy construction process.
\item The development of a hierarchical taxonomy tailored for the software application domain, demonstrating the applicability of our approach in a real-world case;
\item An assessment of Large Language Models (LLMs) for the automatic evaluation of domain-specific taxonomies;
\end{itemize}

We make the code~\cite{cezar_sas_2024_13545534} and data~\cite{sas_2024_13546689} available.

 This paper is structured as follows: Section \ref{sec:background} provides the necessary background and motivations for our study. Section \ref{sec:datasources} describes the datasources used in our research. The proposed approach is outlined in Section \ref{sec:proposed} Section \ref{sec:eval} discusses the evaluation metrics and criteria used to assess the performance of our approach. The results of our experiments are presented in Section \ref{sec:results}, followed by a discussion in Section \ref{sec:discussion}. Finally, we review related work in Section \ref{sec:sota} concludes.

\section{Background}
\label{sec:background}

This section defines terminology and highlights the need for a taxonomy by presenting issues in prior works.

\subsection{Terminology}

This paper employs several key terms: \textbf{classification}, \textbf{categorization}, and \textbf{taxonomy}. While these terms are sometimes used interchangeably, they have distinct meanings in the context of this paper.

\textbf{Classification} refers to the process of assigning objects to predefined categories based on their characteristics, typically using machine learning algorithms. It focuses on the act of grouping and the techniques involved.  For example, in an e-commerce platform, a machine learning algorithm might classify products into categories like ``Electronics", ``Clothing", or ``Home Appliances" based on features such as product descriptions, images, and customer reviews.

\textbf{Categorization} pertains to the system or scheme of categories into which objects are sorted. It represents the framework or structure that organizes categories but does not imply the process of grouping. The categorisation scheme in the same e-commerce platform could include broader categories like ``Men's Fashion", ``Women's Fashion", ``Kids", and ``Accessories". These categories provide the structural framework for organizing products, allowing customers to navigate the platform easily. Still, they do not involve the process of placing a specific product into one of these categories.

Lastly, \textbf{taxonomy} is a specialized type of categorization that explicitly defines hierarchical relationships between categories. In our discussion, the taxonomy employs a \textbf{polyhierarchy}~\cite{hedden2016accidental}, meaning that categories can have multiple parent categories.  For example, within the e-commerce platform, a product like a ``Smartwatch" might belong to both the ``Electronics" category and the ``Accessories" category, reflecting its relevance to multiple areas of the catalogue. This polyhierarchical taxonomy allows for a more flexible and comprehensive organization of products, enabling better search and recommendation functionalities.

In this paper, the terms \textbf{categories}, \textbf{classes}, and \textbf{labels} are used interchangeably to refer to the individual units or elements within these frameworks.

\subsection{Why Creating a New Taxonomy?}
\label{sec:motiv}

Defining the terms used to create the hierarchy (e.g., the `\textit{seed}') is the necessary first step for building a taxonomy. As part of a previous study~\cite{sas2022antipatterns}, we evaluated several categorisations as proposed by various researchers in the past, aiming to assess each as the potential seed of our work. The assessment of these seeds revealed several recurring issues (or anti-patterns), including:
\begin{itemize}
    \item \textbf{Mixed Taxonomies (MT)}, where the set of labels span across different tasks (\eg application domains and programming languages);
    \item \textbf{Single Category (SC)}, where the projects in the dataset are annotated with a single label;
    \item \textbf{Mixed Granularity (MG)}, where the terms in the taxonomy are in hypernym-hyponym relationships, but these are not explicitly modelled (\eg \lbl{Neural Networks} is a subdomain of \lbl{Machine Learning}, yet they appear as independent labels).
\end{itemize}

Furthermore, while having less impact on model performance, other observed antipatterns included:
\begin{itemize}
    \item \textbf{Non-Exhaustive Categories (NE)}, where there are terms that have a common parent, but one of them is missing (e.g., having \lbl{Interpreters} but not \lbl{Compilers});
    \item \textbf{Non-Relevant Categories (NRC)}, which manifest when there are categories too semantically distant from the general concept of the taxonomy (e.g., \lbl{Boardgames} along with \lbl{Compiler}, \lbl{Interpreter}, and \lbl{Database}), artificially boosting model performance;
    \item \textbf{Unnecessarily Joined Categories (UJC)}, where a label combines two terms (e.g., \lbl{Servers and Networking});
    \item \textbf{Sink Category (SKC)}, where a generic label (e.g., \lbl{Frameworks} or \lbl{Libs}) is used as an umbrella for hard-to-annotate projects;
    \item \textbf{Ambiguity (AMB)}, added in~\cite{sas2022gitranking}, for terms not grounded to a knowledge base, making their definitions open to interpretation by users.
\end{itemize}

Table~\ref{tab:pathologies} summarizes the presence of these antipatterns in prior works, the number of classes used, and how the categorization has been constructed, bottom-up (BU), when the source of the terms is from a pre-existing set defined by users in a collective fashion, or top-down (TD) in case the authors define the terms in a centralized way.

\begin{table}[htbp!]
\centering
    {\begin{tabular}{lcccccccccc}
\toprule
\rot{\textbf{Study}}&  \rot{\textbf{Classes}} & \rot{\textbf{MT}} & \rot{\textbf{MG}} & \rot{\textbf{SC}} & \rot{\textbf{NE}} & \rot{\textbf{NRC}} & \rot{\textbf{UJC}} & \rot{\textbf{\textbf{SKC}}}  & \rot{\textbf{AMB}} & \rot{\textbf{Definition}}
        \\
        \midrule
\cite{Kawaguchi2006MUDABlue}   & 6  & \OK & \OK & \OK  & \OK  & \OK  &      &      &    \OK & TD \\
\cite{tian2009lact}   & 6               &     & \OK & \OK  & \OK  &      &      &      &  \OK  & TD  \\
\cite{vasquez2014api}  & 22    &     & \OK &      &      &      &      & \OK  &   \OK & TD  \\
\cite{Borges2016popularity} & 6  &     & \OK & \OK  & \OK  &      &      & \OK  &   \OK & TD  \\
\cite{leclair2018neural} & 75  & \OK & \OK & \OK  &      &      &      & \OK  &   \OK & TD  \\
\cite{altarawy2018lascad}  & 6   &     & \OK & \OK  & \OK  &      &      &      &    \OK & TD \\
\cite{ohashi2019cnn_code}  & 23  &     &     & \OK  & \OK  &      &      &      &    \OK  & TD \\
\cite{sharma2017cataloging} & 22 & \OK &     & \OK  & \OK  &      & \OK  & \OK  &    \OK & TD \\
\cite{soll2017classifyhub} & 5    & \OK &     & \OK  & \OK  & \OK  &      &      &    \OK & TD \\
\cite{zhang2019HiGitClass} & 2-13$\ast$    & \OK &     & \OK  &      &      & \OK     &      &   \OK & TD  \\
\cite{zhang2019HiGitClass} & 3-10$\ast$    & \OK &     & \OK  &      &      & \OK     &      &   \OK  & TD \\

\cite{izadi201repologue}$^\diamond$ & 228  & \textbf{-}$^\dagger$  & \OK &      & \OK  &      &      &      &  \OK$^\dagger$    & BU \\
\cite{sipio2020naive} & 134     & \OK & \OK &      & \OK  &      &      &   & \OK & BU  \\
\cite{sas2022gitranking} & 301  &  & -$^\bigstar$  &      & \OK  &      &      &      &   & BU   \\
\midrule
\multicolumn{11}{l}{$\ast$ Two-level hierarchy. Number of classes for each level.} \\
\multicolumn{11}{l}{$\diamond$ And the extension SED-KG~\cite{izadi_semantically-enhanced_2023}.} \\
\multicolumn{11}{l}{$\dagger$  SED-KG's information allows for filtering and reduces ambiguity.} \\
\multicolumn{11}{l}{$\bigstar$ Has hierarchy but no links between terms.} \\
        \bottomrule
    \end{tabular}}
    \caption{Summary of the statistics of prior taxonomies and the antipatterns. TD refers to `top down', BU to `bottom up'. Adapted from~\cite{sas2022antipatterns} and~\cite{sas2022gitranking}.}
\label{tab:pathologies}
\end{table}

While the \textbf{NE} antipattern is impossible to fully address for very generic taxonomies (\ie in practical cases there will always be new domains to be added to a taxonomy), the \textbf{MG} can be addressed by connecting the terms in a taxonomy based on their \hypohyper relation.

While some taxonomies avoid \textbf{MG} (\eg Ohashi \etal~\cite{ohashi2019cnn_code} and HiGitClass~\cite{zhang2019HiGitClass}), this doesn't mean they are ideal for practical use. Ohashi \etal's labels come from university courses, making them impractical for real-world applications. HiGitClass, though connected, has a narrow focus (\eg AI and Biology). Both cases would benefit from a more general, ground-up approach.


Among the proposed categorizations, the most recent ones \cite{izadi201repologue, sipio2020naive, sas2022gitranking} are all based on labels extracted from GitHub Topics~\footnote{\href{https://github.com/topics}{https://github.com/topics}}, showing a trend towards using real-world labels, instead of imposing them from the top down. While all are sourced from GitHub, GitRanking uniquely selects labels as application domains instead of sampling popular GitHub Topics, removing programming languages and technologies from labels and reducing the \textbf{MT} antipattern.


The next section explains our initial seed categorisation and the datasources used to build the structured hierarchy.

\section{Datasources}
\label{sec:datasources}

We present here the datasources used in the various steps of the taxonomy construction, as well as their characteristics, including advantages and disadvantages. For creating a new seed (\ie the first step of the taxonomy creation), we used GitRanking~\cite{sas2022gitranking}. This automatic classification tool uses the labels attached to existing software projects by GitHub developers. GitRanking offers practical, real-world categories derived from GitHub Topics, reflecting actively used categories.

To create the hierarchical structure, we used three different datasources. The first is the Computer Science Ontology (CSO)~\cite{salatino2018cso}, which provides a structured and comprehensive classification of computer science concepts derived from academic papers. The second source is Wikidata, a collaborative, generic knowledge base constantly updated by a global community, providing a rich and dynamic source of information. Lastly, we also used Large Language Models (LLMs), given their ability to perform well on various natural language tasks, including taxonomy construction.

\subsection{GitRanking}
We use GitRanking~\cite{sas2022gitranking} as our initial seed of terms. GitRanking categorization is made by a curated set of software application domains manually derived from GitHub Topics. This bottom-up approach follows the source of content principles~\cite{hedden2016accidental} of gathering the information from the content (\ie repositories), and the people (\ie developers or library users) as the terms have been collected from the final repositories. The terms have been defined and filtered by users.
GitRanking excludes non-application domains, such as programming languages and technologies. Each term in the list is associated with a corresponding Wikidata entry via its Wikidata ID, allowing for easy disambiguation of the terms.

Although manually curated, the taxonomy lacks defined hierarchical relationships between terms. This omission creates challenges for classification tasks, as most terms in the taxonomy share an `\textit{IS-A}' relationship. Additionally, the annotated data often lacks both hypernym and hyponym labels, further complicating classification due to overlapping features between various labels. For instance, a project labelled with \lbl{Neural Network} should also be labelled with \lbl{Machine Learning}, but this consistency is not always maintained.

Another characteristic of GitRanking is that the terms are sorted from the most generic (\eg \lbl{Science}) to the most specific (\eg \lbl{Convolutional Neural Network}). The sorting was done using manual pairwise comparison, where the user will select the most generic term. 
The TrueSkill~\cite{herbrich2007trueskill} algorithm, used for ranking players in competitions like chess or online games, was then applied. This ranking information will be utilized in the later post-processing step.

\subsection{Computer Science Ontology (CSO)}

The Computer Science Ontology (CSO) is an automatically generated ontology of research topics in the computer science field. It was created using Klink-2~\cite{osborne2015klink2}, a method for automatically generating semantic topic networks by analyzing networks of research entities, such as papers, authors, venues, and technologies, to infer semantic relationships between topics. CSO was developed using 16 million scientific articles in the computer science domain and models various types of entities and their relationships. In this work, we are interested in the following relations: 

\begin{itemize}
    \item \textbf{cso:superTopicOf}: describes the ``IS-A'', or hypernym, relation among the terms in the ontology. It is the main relation we are interested in.
    \item \textbf{cso:preferentialEquivalent}: many of the terms might have different surface forms, aliases, that are used to refer to the same entity. This relation points all of these aliases to a common term. We use this to disambiguate multiple entities and remove any duplicates.
    \item \textbf{owl:sameAs}:  models the relation between entities in the CSO and ones in other knowledge bases (KB), including Wikidata, DBpedia,  Freebase, and more. We will use this as one of the approaches to link the terms in GitRanking to the ones in CSO, as GitRanking terms are disambiguated with Wikidata IDs.
\end{itemize}

While automatically created, CSO has been used in real-world cases (\eg Springer Nature) and refined with user feedback multiple times, with the last update on August 2021.

\subsection{Wikidata}
Wikidata~\cite{vrandecic2012wikidata} is a collaboratively edited knowledge graph. It contains a machine-readable version of information that is available on Wikipedia. It is designed around entities, each representing a topic, concept, or object in the real world and is identified by a unique identifier (QID). 

As for CSO, Wikidata also models the relations between the entities; their relations are called properties, and each has a property ID (PID).

For our task, we are interested in the following properties:

\begin{itemize}
    \item \textbf{subclass of (P279)}: is the ``IS-A'' relationships in Wikidata. For example \lbl{Machine Learning} has a P279 relationship with the entities \lbl{Computer Science} and \lbl{Artificial Intelligence};
    \item \textbf{instance of (P31)}: describes the type of an entity. For example, for \lbl{Machine Learning}, we have that for property P31, there are the following entities: \lbl{academic discipline}, \lbl{certification}, \lbl{field of study}, and more.
\end{itemize}

We will use these two properties to traverse the graph and construct our taxonomy.


\subsection{Large Language Models}
Given their increasing popularity due to their performance in this and various tasks, we also evaluate the ability of LLMs to perform in constructing a taxonomy given our categorization. While they lack domain-specific knowledge, LLMs are trained on large amounts of data, spanning the content of almost the entire web~\cite{brown2020gpt}. Additionally, LLMs have been used in taxonomy construction both as assistants~\cite{joachimiak2024artificial} and for fully autonomous creation~\cite{chen2023llm_taxonomy, gunn2024taxonomy_llm}.

\section{Methodology}
\label{sec:proposed}

\begin{figure*}[htbp]
    \centering
    \includegraphics[width=\linewidth]{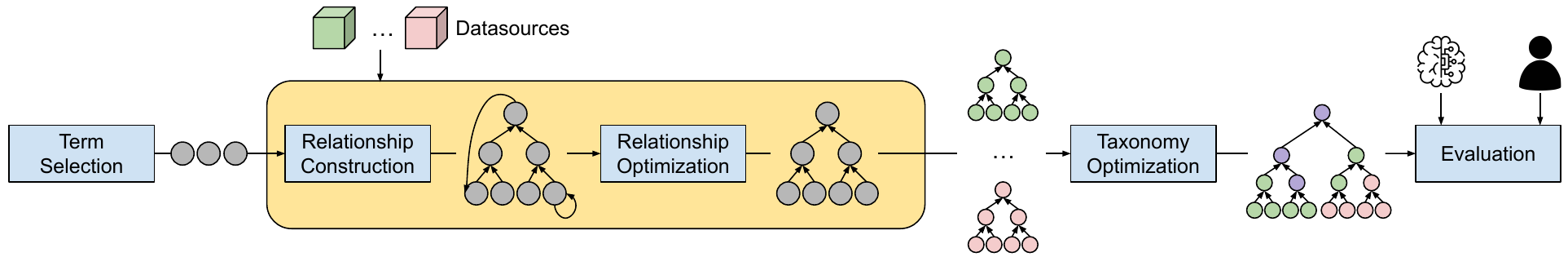}
    \caption{Pipeline: i) select terms, ii) optimise relationships per datasource, iii) combine results, and iv) evaluate.}
    \label{fig:pipeline}
\end{figure*}

For our methodology, we follow the usual taxonomy construction pipeline \textbf{Terms Selection}, \textbf{Relationship Construction}, \textbf{Evaluation}. Moreover, since we work with automatic construction, among these three steps, we incorporate additional steps within this framework: \textbf{Relationship Optimization}. This is achieved in two steps:
\begin{enumerate*}
    \item \textbf{Cleaning}: wrong \textit{IS-A} relationship that can be removed (\ie cycles) will be removed in this step~\cite{wang-etal-2017-short};
    \item \textbf{Best Model Selection}: since for each datasource, we use various hyperparameters, and also cleaning steps, we need to pick the best configuration for the taxonomy construction.
\end{enumerate*}
Lastly, we perform \textbf{Taxonomies Optimization} by performing an ensemble to combine the terms and relationship of multiple taxonomies; resulting in a more complete and usable taxonomy. This step is common when organizations merge and must combine their taxonomies~\cite{hedden2016accidental}. A visual representation of the pipeline is presented in Figure~\ref{fig:pipeline}.

To make our taxonomy well-defined in the scope and intended use and improve reproducibility, we also expressed the construction principles that will guide our choices in defining the algorithms and choices for the construction.

\subsection{Construction Principles}

For the taxonomy's construction, we follow similar principles and guidelines used when manually creating a taxonomy. In particular, we use the principles defined in~\cite{hedden2016accidental}.

\subsubsection{Purpose, Users, Content, Scope}
We first start by answering the following questions:

\begin{description}
    \item \textbf{What will the taxonomy be used for? (Purpose)} 
    
    The purpose of the taxonomy is to map the software application domains and allow for better classification and retrieval of software repositories in codebases.
    
    \item \textbf{Who are all the taxonomy users? (Users)} 
    
    They will be developers, researchers, and domain experts.
    
    \item  \textbf{What content will the taxonomy cover? (Content)} 
    
    The content covered is software application domains.
    
    \item \textbf{What are the topic area, scope, and limits? (Scope)} 
    
    The taxonomy does not cover programming languages or technologies. Furthermore, at the moment, it is targeted toward GitHub Topics that are application domains. The choice is based on GitHub being the most popular repository hosting platform. Furthermore, this choice will also allow for a source of annotated repositories that can be used for training ML models. 
\end{description}

\subsubsection{Principles for inclusion of terms}
Now that we have defined the taxonomy's purpose, users, content, and scope, we also need to define the principles that guide us in constructing it, particularly the inclusion of terms. These principles will help by providing a framework for decision-making, resulting in more consistent choices. The principles for the terms are:

\begin{itemize}
    \item \textbf{Relevance to Subject Area}: Ensure the concept falls within the intended scope of the taxonomy.
    \item \textbf{User Interest}: Prioritize concepts that are important and likely to be sought after by users.
    \item \textbf{Information Availability}: Confirm that sufficient information is available on the concept.
    \item \textbf{User Demand and Expectation}: Address concepts users want and expect to be covered.
\end{itemize}

\subsubsection{Principles for including hierarchical relationships}
For hierarchical relationships, we follow these principles:

\begin{itemize}
    \item \textbf{Inclusion and Scope}: Each narrower term must be included within its broader term, encompassing more than just the members of a single narrower term.
    \item \textbf{Generic-Specific Relationships}: Define only \hypohyper (generic-specific) relationships.
    \item \textbf{Polyhierarchy}: Allow terms to have more than one generic term to accommodate polyhierarchy.
    \item \textbf{No Circular References}: Prevent cycles and loops to avoid circular references.
    \item \textbf{Controlled Depth}: Limit taxonomy depth to avoid overly nested structures and ensure easy navigation.
\end{itemize}

\subsection{Terms Selection}

We start the construction of our taxonomy by selecting an initial seed. We do this by evaluating the existing flat categorization in the literature. We base our selection on the antipatterns described in our previous work and summarized in Section~\ref{sec:motiv}, as we want the seed with the fewest prior issues. From this, we opted for GitRanking as our seed, which contains terms such as \lbl{Machine Learning}, \lbl{Graphical User Interface}, \textit{Database} and more. 


Besides having the least amount of antipatterns, GitRanking is a curated list of GitHub's Topics. Given its bottom-up construction, this ensures that the topics are relevant to the subject (manually curated) and of interest to the user. Furthermore, each topic is linked to a Wikidata entity, giving us enough information about each topic. Bottom-up taxonomies offer significant advantages: by emerging organically from user input and real-world data, they provide high flexibility by quickly evolving with the subject matter~\cite{vet1998bottom}. This approach ensures that taxonomies are accurate and contextually relevant, aligning closely with user interests. Moreover, the integration with Wikidata entities enhances the richness of information, allowing for a better understanding of each topic (addressing the ambiguity antipattern).

\subsection{Relationship Construction}

Starting with our initial list of terms, derived from GitRanking, we integrate information from various datasources to build a comprehensive, structured, hierarchical taxonomy of software application domains.

For this taxonomy construction, we selected three distinct datasources: CSO, Wikidata, and LLMs, each offering unique specificity and characteristics. CSO serves as a domain-specific ontology, Wikidata provides a broad general knowledge base, and LLMs offer both general domain knowledge and generative capabilities.

The process begins with a chosen datasource. For each GitRanking term, we match it to a corresponding term in the source (except for LLMs) and then perform a graph traversal from that point.

\subsubsection{GitRanking-based CSO}

To create the taxonomy structure, we start by matching each GitRanking term with its CSO equivalent: both GitRanking and CSO contain a link to Wikidata entities via their QID. However, we noticed that the original research in~\cite{salatino2018cso} have linked incorrectly many CSO terms to Wikidata QIDs. For example, the GitRanking term \lbl{semi-supervised clustering} is also contained in CSO with the QID \textit{Q1041418}. However, that QID in Wikidata refers to \lbl{semi-supervised learning}.

\begin{figure}[htbp!]
    \centering    \includegraphics[width=.5\textwidth]{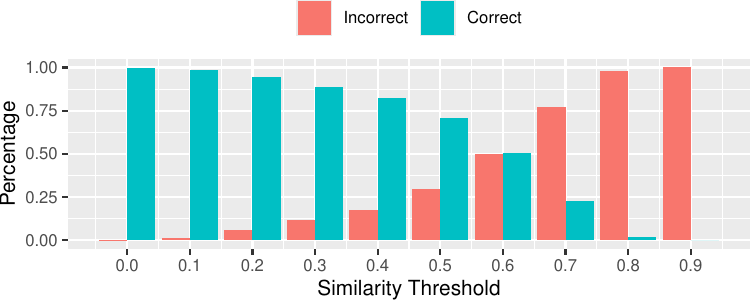}
    \caption{Percentage of correctly linked terms to Wikidata in CSO at various similarity thresholds.}
    \label{fig:wiki_correctness}
\end{figure}

This issue appears more widespread than a few instances: therefore, we systematically evaluated the quality of CSO terms linked to Wikidata using semantic similarities between the linked Wikidata and CSO terms.  In Figure~\ref{fig:wiki_correctness}, we observe that with a conservative similarity threshold of 0.5, 30\% of terms are incorrectly linked, rising to over 50\% at a threshold of 0.6. Since the terms represent the same entity, their similarity should be high. We will consider this finding when selecting the optimal parameters for CSO.

Having identified these errors and the corresponding QIDs, we incorporate a semantic similarity check between the aliases of the Wikidata entities and those of the CSO. For the semantic similarity, we evaluate two embedding models: `MPNet-Base' (all-mpnet-base-v2) and `MiniLM-L6' (all-MiniLM-L6-v2). These models are state-of-the-art models\footnote{\href{https://www.sbert.net/docs/pretrained_models.html}{https://www.sbert.net/docs/pretrained\_models.html}} for representation learning (\ie embedding), with the main difference between them being the number of parameters; the L6 model is smaller. The thresholds used vary from 0.0 to 1.0, with a step of 0.10

After pairing the GitRanking terms with their corresponding CSO entities, we traverse the graph using the \textbf{cso:superTopicOf} link. The traversal is how we find each term's parents and ancestors. Finally, we eliminate redundant edges for synonymous terms adopting the \textbf{cso:preferentialEquivalent} link.

\subsubsection{Wikidata based on GitRanking}
To construct the taxonomy using Wikidata, we use the API provided. For each term in GitRanking, each having a QID, we query the API to get the Wikidata entity. In the entity, we check all the other entities in \textbf{subclass of (P279)} relation with the current one and add them to our stack. We iterate this process for all the newly found terms and the remaining ones in GitRanking.

One issue we encounter with using Wikidata is the presence of very generic and abstract terms (\eg \lbl{intentional human activity}, \lbl{temporal entity}, \lbl{action}). We evaluate various strategies to reduce the issue during construction.

Aside from evaluating the construction without any condition (TA=T), we use a max depth limit (MD), so when traversing the ancestors, we limit the maximum number of edges between the term and the further ancestor (we evaluate multiple depths). Furthermore, using the property \textbf{P31}, an instance of Wikidata, we can limit the terms to an instance of terms also present in the hand-picked GitRanking list of terms. We use a threshold on the frequency of the types (TT).

\subsubsection{LLMs  based on GitRanking}

We opt for GPT3.5 (gpt-3.5-turbo) and GPT4 (gpt-4-1106-preview) as the models we will use for the construction. Besides the model, when using LLMs, the choice of the structure of the prompt is also important. Following the prior work~\cite{gunn2024taxonomy_llm}, we use two prompts for our taxonomy construction, one containing the list of terms (WT) and the other without (NT), with just the term of interest. However, unlike \cite{gunn2024taxonomy_llm}, we do not use an initial taxonomy structure, which will require human intervention to construct.

For the first case, the prompt template is shown in LLM Prompt~\ref{llm_1}, while for the second, it is shown in the Appendix.
The angular brackets are placeholders for the terms in the taxonomy and the term of interest.

\begin{llmprompt}[label=llm_1]
\justifying
\texttt{
You are a helpful assistant tasked to pair terms to the hypernym to which they should belong. If it does not belong to any, answer None. \\ 
Given a term, provide the hypernym for the term. \\
Multiple answers are allowed, and should be separated by a comma. Keep the answer concise, in CSV format, without any extra. \\
For example: \\
parent1 \\
parent1,parent2,parent3 \\
None \\
What is the hypernym of \{term\}?}
\end{llmprompt}


\subsubsection{Iterative LLMs based on GitRanking}
While prior research shows promising results~\cite{joachimiak2024artificial,chen2023llm_taxonomy}, the GPT-4 generated results also suffer from excessive granularity, resulting in heavy repetition within similar but different categories~\cite{gunn2024taxonomy_llm}. To manage this issue, we also evaluate an iterative approach (LLM\_Iter), where we try to build upon each newly added term. In this case, we use the same prompt templates.

\subsection{Relationship Optimization}

Once we create the taxonomy, we need to remove edges (and nodes) that do not add information and, therefore, only increase its complexity. Furthermore, we need to identify which configurations of hyperparameters work best for each datasource, hence producing the best relationships. Lastly, we improve the relationships by combining the relationships (and nodes) from multiple taxonomies.

\subsubsection{Cleaning}

The resulting taxonomy may have issues like irrelevant terms or loops. Irrelevant terms clutter the taxonomy, while loops create confusion in the hierarchy. We use post-processing to remove cycles and abstract edges, refining the structure. Each step is detailed below.



\textbf{Cycle Removal:} The resulting taxonomies might not direct acyclic graphs (DAGs); this makes the taxonomy unusable as a cycle and will result in the inability to traverse the graph to annotate examples with the hierarchical structure. Furthermore, some self-loops might also appear.

We start by removing all self-loops. For larger cycles, we use heuristics: if a node’s parent is at a greater depth from the root of the subtree, we remove the edge. For example, some pairs that are removed are \relation{bioinformatics}{genomics}, \relation{task}{activity}, and \relation{cryptography}{identity-based}.


\textbf{Abstract Edges Removal:} One of the main issues with the Wikidata-built taxonomy is the large number of edges created by generic terms, which results in an overly complex structure. To address this problem, we implement a strategy focused on edge removal by limiting the number of connections to ancestors for certain nodes.

In particular, we target the most generic terms for this edge-pruning process. In GitRanking, terms are organized into clusters ranging from the most generic to the most specific. We remove parent edges from the most generic terms in GitRanking, assuming that unnecessary terms are likely among their ancestors. This approach simplifies the graph by reducing the number of edges and nodes. Examples of terms removed are shown in Table~\ref{tab:abstract_removed}.

\begin{table}[htbp!]
    \centering
    \caption{Examples of abstract terms removed from a taxonomy. Note that for GPT4, these are also affected by the prompt; in this case, we put a bad case scenario.}
    \label{tab:abstract_removed}
    \begin{tabular}{c|c|c}
    \toprule
         \textbf{CSO} & \textbf{Wikidata} & \textbf{GPT4} \\
         \midrule
         - & abstract class  & abstract idea \\
         - & process  & concept \\
         - & knowledge  &  mammal \\
         - & temporal entity  & notion \\
         - & work  & worker \\
         \bottomrule
    \end{tabular}
\end{table}

\subsubsection{Best Model Selection}

Given the large number of taxonomies constructed from each datasource, due to varying hyperparameter configurations, selecting the best one is essential. Since no ground truth is available and multiple metrics need to be evaluated, the selection process becomes non-trivial.

One critical metric for this selection is the \textit{number of unlinked terms}. The motivation behind this choice lies in the principle of inclusion and the need to meet the users' interests and demands and expectations. Many unlinked terms suggest gaps in the hierarchy, reducing the taxonomy's completeness and usability. By minimizing unlinked terms, we aim to ensure that all relevant concepts are well-represented and connected, increasing the likelihood of producing a usable taxonomy. 

However, this metric cannot stand alone: other factors, such as the density of the taxonomy, its depth, and more, must also be considered to maintain the accuracy and relevance of the taxonomy. The metrics used are presented in Section~\ref{sec:eval}.

We perform the selection using the Technique for Order of Preference by Similarity to Ideal Solution (TOPSIS)~\cite{tzeng2011multiple}. This multi-criteria decision analysis method evaluates alternatives based on their geometric distance from an ideal solution. We opted for TOPSIS because it is a well-established method with an easily interpretable solution. 

The algorithm begins by constructing a decision matrix, where each entry represents the performance of an alternative with respect to a criterion. The matrix is then normalized to account for different scales. Next, a weighted normalized decision matrix is created, applying equal weights to all criteria. Positive-ideal and negative-ideal solutions are determined based on benefit and cost criteria. The distance of each alternative from these ideal solutions is calculated, followed by computing the similarity of each alternative to the ideal solution. Finally, alternatives are ranked in descending order of similarity, with higher values indicating better performance.

Lastly, we check if the solution lies on the Pareto front.

\subsection{Taxonomy Optimization}
Given the different characteristics of each datasource, we integrate and merge~\cite{hedden2016accidental} the different taxonomies into a single one. The integration allows taxonomies that cover various concepts to be combined into one while merging joins similar terms into a singular concept.

We evaluate two different ensemble methods. The first is a simple \textbf{union} of all terms and pairs into a single taxonomy. Then, a semantic similarity analysis is performed to find duplicate terms with minor surface form differences.

The second approach is a \textbf{cascade} method, building the taxonomy by adding GitRanking term paths not linked in the first step, but identified in the second.

Finally, we use an LLM to link any remaining unconnected terms. The same prompt used for taxonomy construction is applied here.

\section{Evaluation}
\label{sec:eval}


Given the lack of ground truth, subjectivity, and multiple optimal solutions, an automatic and objective evaluation of taxonomy quality is impossible. Instead, we use a combination of automatic (metrics and LLM) and human-based (annotators) evaluations. For the automatic evaluation (\ref{sec:auto_evalu}), metrics help find the best model and understand each source's characteristics. Human-based (\ref{sec:human_evalu}) and LLM-based (\ref{sec:llm_evalu}) evaluations were performed only on the best taxonomy due to the high time cost of reviewing all pairs.

\subsection{Automatic Evaluation}
\label{sec:auto_evalu}


The previous indicators and principles are general and subjective. We need to find objective metrics that quantify these aspects, aid to better understand the constructed taxonomies, and select the best model for each datasource.

While several metrics have been proposed to evaluate taxonomy structures~\cite{kaplan2022evaluation, Unterkalmsteiner2023evaluation}, they typically assume a strict hierarchical structure without polyhierarchy, which limits their applicability to our case. Therefore, we utilize a set of graph-based metrics that are better suited for evaluating taxonomies with polyhierarchical structures.


We present the metrics used in this study, briefly defining and describing each. Additionally, we outline specific objectives for each metric to guide the optimization phase.

\textbf{\# Nodes}: represents the total number of nodes in the taxonomy, excluding unlinked terms. This indicates the final size of the taxonomy. Fewer nodes are generally preferred to ensure less complexity. However, more nodes may be needed if key domains are missing from GitRanking. We don't use this metric for the optimization.

\textbf{\# New Nodes}: is the number of new terms that are added to the taxonomy but are not in GitRanking. A lower number indicates that the taxonomy remains closer to the seed, maintaining lower complexity, while a higher number adds terms, increasing complexity and the risk of including irrelevant terms. The objective is to minimize.

\textbf{\# Unlinked}: indicates the number of terms in GitRanking that have not been linked to other terms within the taxonomy. A higher count suggests the taxonomy is less representative of the domain, while a lower count implies better coverage of relevant domains, with more linked concepts users expect. The objective is to minimize.

\textbf{\# Edges}: represents the total number of relationships between terms in the taxonomy. A higher number means more relations are modeled, but this depends on the number of nodes. We don't optimize it.

\textbf{Density}: is the ratio of actual links to possible links in the graph. A value of 1 indicates a fully connected graph, while 0 means no edges exist. In this context, lower values are preferred. While polyhierarchy is allowed, too many edges can make the taxonomy less usable. The objective is to minimize.

\textbf{\# Roots}: number of nodes with no parents but at least one child. Ideally, this should be close to 1, meaning all terms fall under one general term. A higher value indicates multiple general domains or potentially many disconnected subgraphs. We minimize it.

\textbf{\# Leaves}: represents the number of nodes with no children but at least one parent. A higher number of leaves indicates that the more specific terms from the initial seeds are linked, while the more general ones are not. Combined with a low diameter, this can suggest that the taxonomy is shallow and wide. We don't use this metric for the optimization.

\textbf{Max and Avg Parents}:  tracks statistics regarding number of parents across the taxonomy. A high number of parents suggests that one term intersects with many subdomains. We don't optimize them.

\textbf{Avg Depth}: is the taxonomy's average distance between root and leaf nodes. Higher values indicate deeper information structures, but can also be influenced by a single deep path. Low values may suggest a shallow taxonomy, potentially linking terms to overly generic parents. We don't use this metric for the optimization.

\textbf{Diameter}: is the maximum shortest path between nodes in the graph. It indicates the overall depth of the taxonomy. A higher diameter suggests that one branch may have overgrown, making the taxonomy less usable, while a very low value implies poor connectivity. However, a too low bound is not reached in our case. The objective is to minimize.

\textbf{Max and Avg Children}: tracks both the number of children for nodes in the taxonomy. A node with many children likely represents a domain with numerous subdomains. These metrics are only used for statistical purposes. We don't optimize them.

\textbf{\# Components}: he number of weakly connected components in the taxonomy. A higher number of components suggests the taxonomy includes multiple subdomains that are not linked, while a lower number indicates better connectivity. The objective is to minimize.

\textbf{\# Loops}: refers to the number of edges that link a node to itself. The ideal value for this metric is zero, as no term should be its parent, which would create circular references and render the taxonomy unusable. We don't optimize it.

\textbf{\# Cycles}: indicates the of walks that begin and end at the same node. They are undesirable because circular references make the taxonomy unusable. We minimize it.

\subsection{Human Evaluation}
\label{sec:human_evalu}

We perform a human evaluation of the best-resulting taxonomy. We use 7 annotators with diverse backgrounds and expertise in computer science, software engineering, artificial intelligence, biology, and biomedical engineering. Their roles span from PhD students to industry developers.

The task is framed as binary annotation; for a given pair \hypohyper, they have to mark it with \textit{0}, wrong pair, or \textit{1}, the terms are correctly linked. Annotators are instructed to annotate only familiar terms to avoid guessing. 

We annotate each pair with at least two annotators; we measure the inter-rater reliability using Krippendorff's alpha~\cite{krippendorff1970estimating, krippendorff2004reliability}, a more general and reliable measure for inter-rater reliability compared to Cohen's Kappa~\cite{cohen1960coefficient}, suited for any number of annotators and incomplete data.

\subsection{LLM Evaluation}
\label{sec:llm_evalu}

LLMs are becoming increasingly popular, given their ability to perform well for various tasks. They have been used as human evaluation alternatives for various tasks, showing close to human performance~\cite{chiang-lee-2023-large, chiang-lee-2023-closer}, or at least can be a good starting point~\cite{lu2023human}.
Therefore, we prompt ChatGPT to evaluate the quality of the resulting taxonomies by checking each pair. For the prompt, we follow~\cite{gao2023rate} (see Appendix). 

To see how the LLM stacks up against the human annotators, we also perform an inter-rater agreement against all the other annotators. This aims to gauge how closely the LLM's results align with human judgments and identify its effectiveness for automatic domain-specific taxonomy evaluation.

\section{Results}
\label{sec:results}


We present the results of taxonomies created from various data sources. Based on hyperparameters and post-processing, we identify the best model for each. Finally, we combine the selected taxonomies in an ensemble model to overcome the individual limitations and show the evaluation results.

\subsection{Taxonomies}

\begin{figure*}[htbp!]
    \centering
    \includegraphics[width=.9\textwidth]{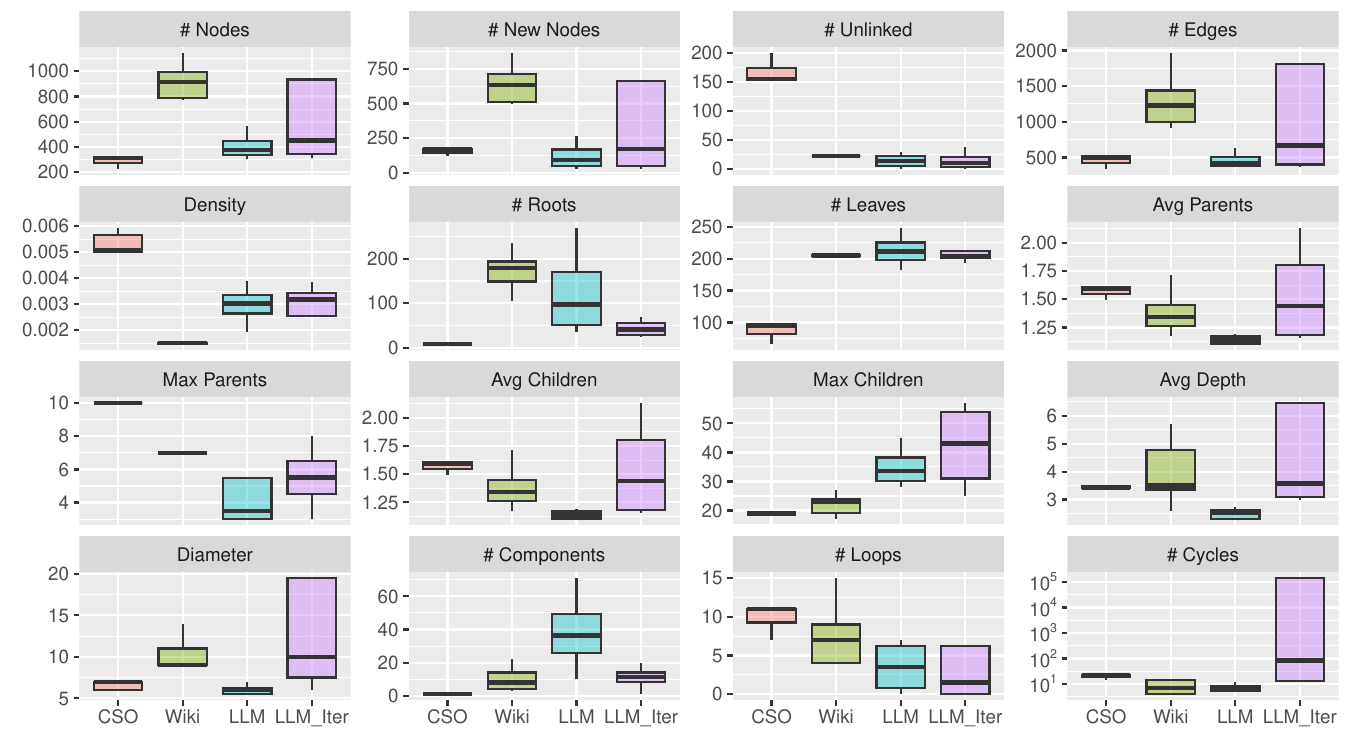}
    \caption{Results for each datasource include a distribution based on various hyperparameter configurations.}
    \label{fig:models_metric_distr}
\end{figure*}

We first present the characteristics of each datasource using the metrics illustrated in Figure~\ref{fig:models_metric_distr}, followed by the results of the best model selection and a summary. 

\subsubsection{Terms Selection}

Examining the number of new terms (\ie \textit{\# New Nodes}), CSO has a relatively low count, and its total number of terms (\ie \textit{\# Nodes}) is among the lowest compared to other models. However, the \textit{\# Unlinked} metric is the highest for CSO. This can be attributed to its construction from scientific papers, which might not cover some technical or newer terms. Additionally, the problem of CSO with incorrectly linked terms to Wikidata increases this count by approximately 10 terms. This suggests that while the more domain-specific datasource minimizes the introduction of irrelevant terms in the taxonomy, it also omits many relevant terms due to their specificity and not fully overlapping with the taxonomy's intended usage.

The Wikidata-based taxonomies are considerably larger, with the total number of nodes around 1100. This indicates that while Wikidata adds many new terms, it links many more than CSO. However, the \textit{\# Unlinked} is very low and with low variance. However, the \textit{\# New Nodes} has some variance, indicating that we might add excessive nodes in some cases.

Moving to the simple LLM-generated taxonomies, we can see they showcase a mix of variances across different metrics. Specifically, they have the lowest number of new terms, both in \textit{\# Nodes}, \textit{\# New Nodes}, and \textit{\# Unlinked}, suggesting adequate coverage of GitRanking terms without introducing unnecessary or irrelevant terms. However, the taxonomies also feature a relatively high number of \textit{Roots}, which indicates a lack of hierarchical structure.

Lastly, we want to point out that the LLM\_Iter model exhibits exceptionally high variance, making it unsuitable for high-quality and reliable results. Therefore, we have opted to exclude it from further experiments and avoided discussing it.

\begin{boxH}
CSO adds fewer new terms and links, avoiding irrelevant ones but possibly missing important terms. Wikidata and LLMs link nearly all terms but include more noise. The `Term Selection` phase reveals a trade-off between relevance and coverage.
\end{boxH}

\subsubsection{Relationships Construction}

For the edges, CSO has the highest \textit{Density} among all sources, coupled with a high \textit{Avg Parents} and the highest \textit{Max Parents} count. This implies a dense network of related terms or the presence of a very large cluster, the main component, as CSO consists of only one component. The \textit{Avg Depth} and \textit{Diameter} metrics are within a range that balances descriptive detail with usability\footnote{This is relative to the other datasources, as there is no established ground truth or optimal value for these metrics}.

Wikidata-generated taxonomies structure have many \textit{\# Edges} but a very low \textit{Density}. Coupled with a high number of \textit{\# Roots} and \textit{\# Components}, this suggests the presence of many sparse sub-taxonomies. Additionally, the produced taxonomies have large \textit{Avg Depth} and \textit{Diameter} metrics, that with the low \textit{Density} indicate chain-like structures. This is also supported by the lower \textit{Avg Children} metric. As for CSO, the presence of \textit{\# Loops} and \textit{\#Cycles} in the taxonomies generated using Wikidata can be addressed with post-processing.

Concerning the \textit{\# Edges}, the simple LLM-generated taxonomies do not contain many edges, but they exhibit better \textit{Density} compared to those built using Wikidata. This implies a more efficient connectivity among the included terms. However, despite covering a wide range of terms, the structural integrity of these taxonomies is compromised. They contain many \textit{Components} and are relatively shallow, with an \textit{Avg Depth} of just 2. This shallow depth indicates that many disconnected subgraphs are linked to a single node, as evidenced by a high \textit{Max Children}. This suggests that while the model covers a broad spectrum of terms, it fails to establish meaningful connections, leading to a fragmented structure.

\begin{boxH}
CSO creates cohesive but shallow taxonomies with all terms connected. Wikidata taxonomies are deeper but more fragmented. LLM-generated taxonomies cover many terms but are shallow and poorly connected. \\
The `relationship contruction' phase reveals a trade-off between depth, connectivity, and structure.
\end{boxH}

\subsubsection{Relationship Optimization}

Moving to pick the optimal hyperparameter configuration for generating the taxonomies for the datasources, Table~\ref{tab:hyper_cso} presents the results for CSO. For the top 10, we can see that their score is around 0.77, very close to the ideal result of 1 (\ie the solution found is the optimal solution across all metrics). Concerning the configuration, they all have a high similarity threshold (Sim Thresh) score, aligning with the idea that low similarity adds too many irrelevant terms. There is no difference in the embedding model (Emb) used, at least for the high similarity levels. Lastly, we can see that the cycle removal is necessary, while the abstract does not affect it. For CSO there is no abstract term, so this is expected. The final best configuration for CSO is the one with rank 1 in Table~\ref{tab:hyper_cso}.

\begin{table}[htbp]
\centering
\caption{Top 10 hyperparameter configurations for CSO, sorted by TOPSIS score, with model values (LLM), similarity threshold, and postprocessing (Cycle and Abstract).}
\label{tab:hyper_cso}
\begin{tabular}{clcccc}
\toprule
\textbf{Rank} & \textbf{Emb} & \textbf{Sim Thresh} & \textbf{Cycle} & \textbf{Abstract} & \textbf{Score} \\
\midrule
1 & MiniLM-L6 & 0.80 & 1 & 0 & 0.77 \\
2 & MiniLM-L6 & 0.80 & 1 & 1 & 0.77 \\
3 & MPNet-Base & 0.70 & 1 & 0 & 0.77 \\
4 & MPNet-Base & 0.70 & 1 & 1 & 0.77 \\
5 & MPNet-Base & 0.80 & 1 & 1 & 0.77 \\
6 & MPNet-Base & 0.80 & 1 & 0 & 0.77 \\
7 & MiniLM-L6 & 0.90 & 1 & 1 & 0.76 \\
8 & MiniLM-L6 & 0.90 & 1 & 0 & 0.76 \\
9 & MPNet-Base & 0.90 & 1 & 1 & 0.76 \\
10 & MPNet-Base & 0.90 & 1 & 0 & 0.76 \\
\bottomrule
\end{tabular}
\end{table}

Regarding the hyperparameter selection for Wikidata (Table~\ref{tab:hyper_wikidata}). As CSO, the score is close, while not as close, to the ideal optimal solution, with a minimal difference for the 10 best-ranked configurations. For the impact of the parameters, setting the TA parameter to False and an MD value of 3 yields the best results. These settings indicate a need to limit the addition of terms and edges to the taxonomies.  This selection in the terms should be done during the construction phase and not after, as indicated by the lack of TA=T (Take All terms) in the 10 best models. Furthermore, the TT parameter appears too restrictive, and it is optimal to use no threshold. Lastly, we can see that, as for CSO, the cycle post-processing is needed, but abstract removal is not required, at least for the ones that already have limits imposed during the construction.

\begin{table}[htbp]
     \caption{Top 10 hyperparameter configurations for Wikidata: TA (Take All), TT (Type Threshold), MD (Max Depth), postprocessing (Cycle and Abstract), with TOPSIS scores.}
    \label{tab:hyper_wikidata}

    \centering
\begin{tabular}{ccccccc}
\toprule
\textbf{Rank} & \textbf{TA} & \textbf{TT} & \textbf{MD} & \textbf{Cycle} & \textbf{Abstract} & \textbf{Score} \\
\midrule
1 & F & 0 & 3 & 1 & 1 & 0.67 \\
2 & F & 3 & 3 & 1 & 0 & 0.67 \\
3 & F & 3 & 3 & 1 & 1 & 0.67 \\
4 & F & 5 & 3 & 1 & 0 & 0.67 \\
5 & F & 10 & 3 & 1 & 0 & 0.67 \\
6 & F & 5 & 3 & 1 & 1 & 0.66 \\
7 & F & 10 & 3 & 1 & 1 & 0.66 \\
8 & F & 0 & 4 & 1 & 1 & 0.66 \\
9 & F & 0 & 3 & 1 & 0 & 0.66 \\
10 & F & 5 & 4 & 1 & 1 & 0.65 \\
\bottomrule
\end{tabular}
\end{table}

Lastly, for the LLM, in our analysis of the hyperparameter configurations, as shown in Table~\ref{tab:hyper_llm}, we found that GPT-4 consistently outperforms GPT-3.5, with the WT prompts yielding better results than the NT prompts. This suggests that more detailed and context-aware prompts enhance taxonomy generation. As for the previous two, cycle removal post-processing is necessary for improving the structure, whereas abstract post-processing is not required. 

\begin{table}[htbp]
\caption{Top 10 hyperparameter configurations for LLM, including LLM model and Prompt values, and postprocessing.}
    \label{tab:hyper_llm}
    \centering
\begin{tabular}{clcccc}
\toprule
\textbf{Rank} & \textbf{LLM} & \textbf{Prompt} & \textbf{Cycle} & \textbf{Abstract} & \textbf{Score} \\
\midrule
1 & GPT4 & WT & 1 & 0 & 0.65 \\
2 & GPT4 & WT & 1 & 1 & 0.65 \\
3 & GPT3.5 & WT & 1 & 0 & 0.59 \\
4 & GPT3.5 & WT & 1 & 1 & 0.59 \\
5 & GPT4 & WT & 0 & 0 & 0.57 \\
6 & GPT4 & WT & 0 & 1 & 0.57 \\
7 & GPT3.5 & NT & 1 & 0 & 0.57 \\
8 & GPT3.5 & NT & 1 & 1 & 0.57 \\
9 & GPT4 & NT & 1 & 0 & 0.50 \\
10 & GPT4 & NT & 1 & 1 & 0.50 \\
\bottomrule
\end{tabular}
    
\end{table}

\begin{boxH}
CSO and Wikidata taxonomies work better when noisy terms are filtered \textbf{during} construction rather than relying on \textbf{post-processing}. 
For LLMs, detailed, context-rich prompts (WT) produce higher-quality outputs compared to simple prompts (NT). Removing cycles is crucial for structure in all models while removing abstract terms has no impact.
\end{boxH}


\subsection{Taxonomy Optimization}

After analyzing each datasource, we found that each has different strengths and weaknesses, indicating that an ensemble approach might be beneficial. However, to see whether the ensemble could be helpful, we first need to see whether the models behave similarly or differently on the terms and pairs created and unlinked terms.

\begin{figure}[htbp]
    \centering
    \includegraphics[width=\linewidth]{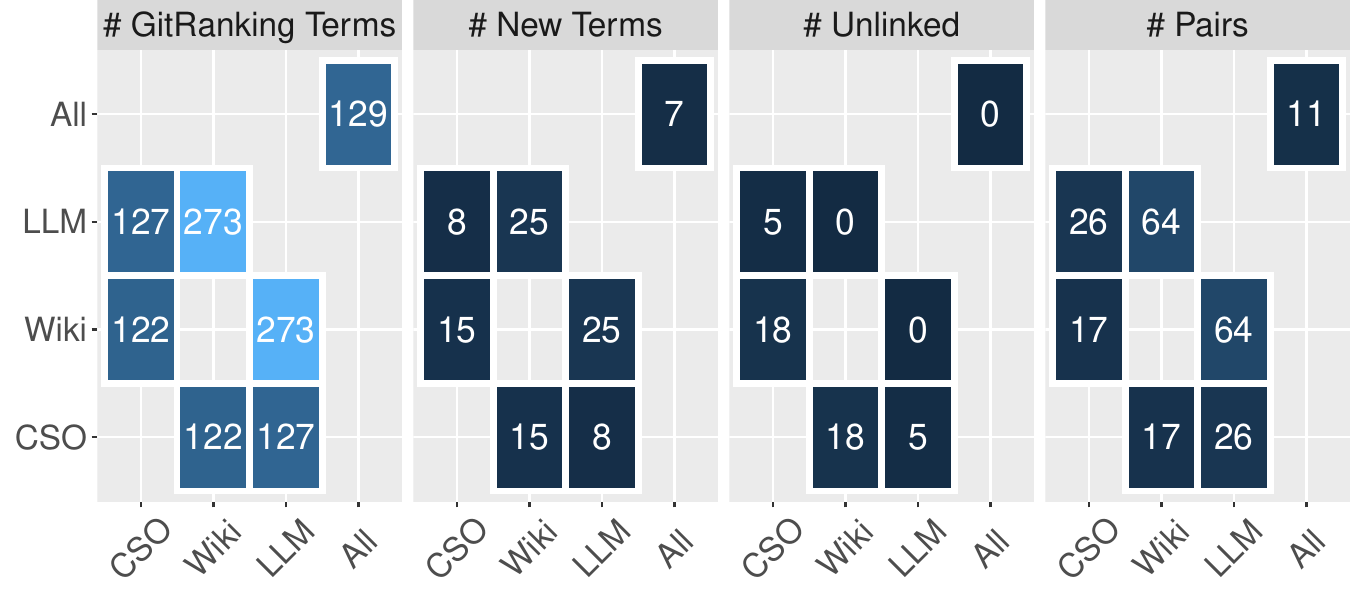}
    \caption{Intersection between models for different metrics.}
    \label{fig:intra_intersections}
\end{figure}

Figure~\ref{fig:intra_intersections} show the number of terms and pairs shared across the different taxonomies. The results show a not surprising behaviour from the LLM; it has a significant similarity regarding pairs with Wikidata. This is expected as part of the training set for GPT is Wikipedia~\cite{brown2020gpt}, the human-readable version of Wikidata. This overlapping suggests that using LLMs as a starting point for our taxonomy results in a similar output as Wikidata, with the added uncertainty of the LLMs generation. Therefore, we opted to keep LLMs only as a completion method for the remaining unlinked terms.

\begin{figure}[htbp!]
    \centering    \includegraphics[width=\linewidth]{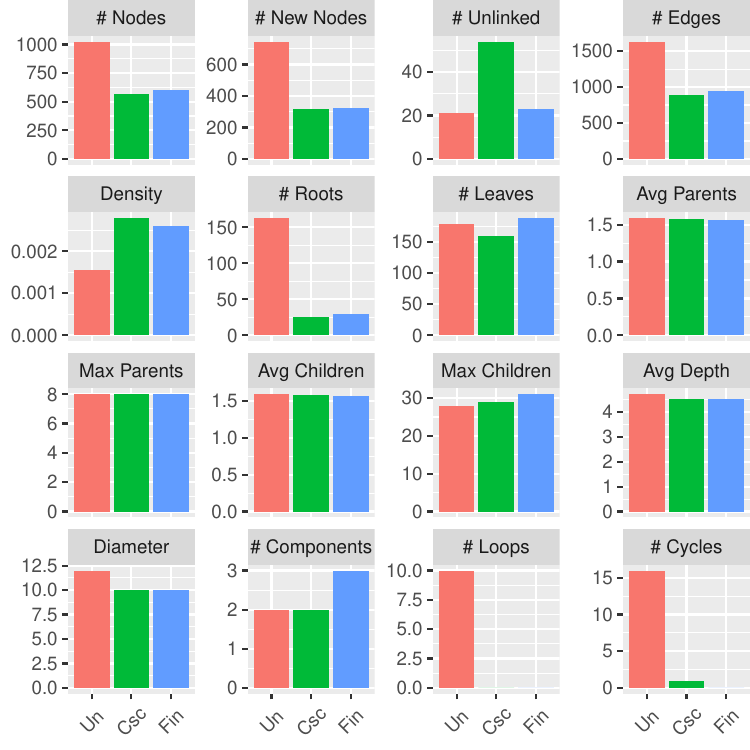}
    \caption{Results for the ensemble models union (Un), cascade (Csc) and the final taxonomy (Fin).}
    \label{fig:ensemble_metrics}
\end{figure}

Shifting focus to the ensemble results, as depicted in Figure~\ref{fig:ensemble_metrics}, we observe that the \textit{\# Nodes} \textit{\# New Nodes}, \textit{\# Edges}, \textit{\# Roots}, and \textit{Diameter} are significantly more, in some cases up to two times more for the union (Un) ensemble compared to the cascade (Csc). This results in Un appraoch creating a more complex taxonomy. While the \textit{\# Unlinked} for Un is half of the Csc, we opt for picking the Csc as it has less \textit{\# New Nodes}. This is based on the principles defined above: the more terms added, the less likely they are to be relevant; complexity and depth are added, which reduces their usability; and lastly, we can link the last missing terms with the LLM.

Having selected the optimal first stage of the ensemble (cascade), we now utilize LLMs to link the remaining missing terms. In Figure~\ref{fig:ensemble_metrics}, we can see the statistics of the final (Fin) taxonomy after the LLM. Notably, the final taxonomy exhibits a slight increase in \textit{\# Node} and \textit{\# Edges} while a significant decrease of \textit{\# Unlinked}, indicating a greater coverage of the seed's terms while still maintaining a reasonable amount of \textit{\# Components} (increased from 2 to 3). Additionally, the final taxonomy has a lower \textit{Avg Depth} and \textit{Diameter}, suggesting a shallower and more compact hierarchy, which can benefit users as it facilitates more straightforward navigation and understanding. 
However, while the final taxonomy reduces the \textit{\# Unlinked} compared to Csc, it still has 21 terms that are not linked.  Lastly, using LLMs as the final part of the construction shows the better ability of LLMs to add to taxonomies rather than generate from the ground up.

\begin{boxH}
The cascade (Csc) ensemble is the optimal choice due to its lower complexity and fewer \textit{\# New Nodes}. Furthermore, LLMs linked the remaining unlinked terms, resulting in a taxonomy with balanced term coverage, semantic richness, and a user-friendly structure.
\end{boxH}


\subsection{Human Evaluation}
After developing the final taxonomy through our ensemble approach, we conducted a human evaluation to assess the quality of the produced \hypohyper pairs.

Table~\ref{tab:annotator_stats} presents the number of pairs each annotator evaluated and the distribution of scores they assigned (either 0 or 1). Most annotators evaluated around 350 pairs, except for Annotator 1, who evaluated approximately 700 pairs. On average, human annotators assigned a score of 0 to about 30\% of the reviewed pairs.


\begin{table}[ht]
\centering
\caption{Number of pairs considered correct (1) and incorrect (1) for each annotator.}
\begin{tabular}{lcccccccccccc}
\toprule
\textbf{Annotator} & \textbf{1} & \textbf{2} & \textbf{3} & \textbf{4} & \textbf{5} & \textbf{6} & \textbf{7} & \textbf{GPT} \\
\midrule
\textbf{\# 0} & 210 & 98 & 154 & 110 & 139 & 138 & 132 & 426 \\
\textbf{\# 1} & 486 & 250 & 194 & 238 & 209 & 210 & 219 & 503 \\
\midrule
\textbf{Total} & 696 & 348 & 348 & 348 & 348 & 348 & 351 & 929 \\
\bottomrule
\end{tabular}
\label{tab:annotator_stats}
\end{table}

When examining the accuracy of the predicted pairs, Table~\ref{tab:annot_stats} illustrates the number of pairs and the corresponding number of votes received. The voting reflects the consensus, or lack thereof, among annotators on the correctness of these term-hypernym pairs. The figure reveals that 161 pairs (17.73\%) are not deemed correct by any annotator. Similarly, 16.74\% of the pairs are considered correct by only one annotator and 19.82\% by two annotators. The majority, 396 pairs (43.61\%), are marked as correct by three annotators. Additionally, 19 pairs are presented to more than three annotators, all of whom considered them correct. Overall, 65.44\% of pairs are deemed correct by at least two annotators, indicating the effectiveness of the proposed approach.


\begin{table}[htbp]
    \caption{Number of pairs that have been considered correct by 0 (incorrect), 1, 2, 3 or more than three annotators.}
    \label{tab:annot_stats}
    \centering
    \begin{tabular}{lccccc}
    \toprule
      \textbf{\# Annotators } & \textbf{0} & \textbf{1} & \textbf{2} & \textbf{3}& $\mathbf{>3}$  \\
     \textbf{\# Pairs}   &   161 & 152 & 180 & 396 & 19  \\
      \bottomrule
    \end{tabular}

\end{table}

\begin{figure}[htbp!]
    \centering
    \includegraphics[width=.8\linewidth]{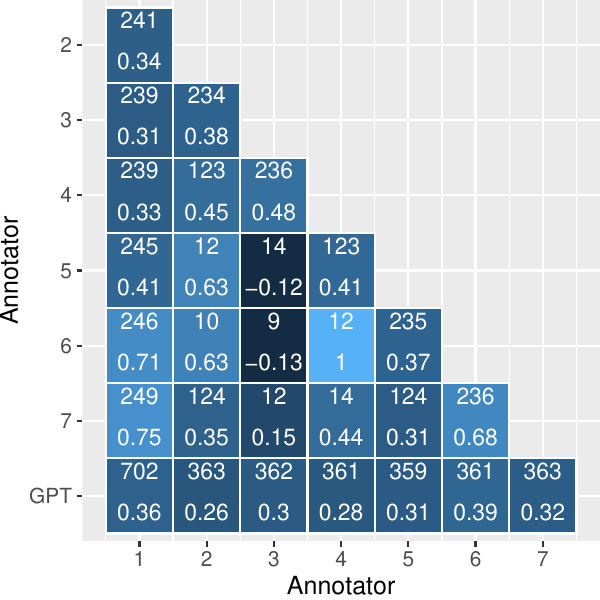}
    \caption{Inter-annotator agreement: the top text shows the number of shared pairs, and the bottom is the agreement score.}
    \label{fig:annot_agreement}
\end{figure}

Finally, we assess the agreement between annotators. Across the entire dataset, Krippendorff's Alpha for all human annotators is 0.46, indicating moderate agreement. The moderate level of agreement underscores the difficulty annotators face in reaching consensus, even when evaluating pre-selected pairs. Figure~\ref{fig:annot_agreement} illustrates the inter-annotator agreement between pairs of annotators. Most agreement scores range between 35 and 50, with higher agreement observed in cases likely due to easier pairs (\eg labels related to machine learning).

The examples in Table~\ref{tab:example_pairs_votes} illustrate the pairs of terms and their associated hypernyms, along with the number of votes they received from annotators. By definition, pairs that received 0 votes or 3 or more votes are generally easier to classify because there is a clear agreement among annotators that these pairs are either correct or incorrect. However, the pairs that received 1 or 2 votes fall into a more ambiguous category, where the terms are less straightforward and more open to interpretation. For example, \lbl{Software project} and \lbl{Software design} (1 vote) or \lbl{Email} and \lbl{Letter} (2 votes) reflect more ambiguity. These pairs might have characteristics that make them seem related, but the connection is not strong or their relevance to the taxonomy might not be obvious. For example, while a \lbl{Software project} involves \lbl{Software design}, the latter term might be seen as a broader concept, making the hypernymy relationship debatable.For \lbl{Email} and \lbl{Letter}, a reviewer might not consider \lbl{Letter} relevant or see the digital version as a hypernym of the physical term.

\begin{table}[htbp]
\caption{Examples of pairs for each amount of votes.}
\label{tab:example_pairs_votes}
    \centering
        \begin{tabular}{lll}
        \toprule
        \textbf{Term} & \textbf{Hypernym} & \textbf{Votes} \\
        \midrule
        Technical standard & Server & 0 \\
        Dimensionality reduction & Discriminant analysis & 0 \\
        Data mapping & Machine-readable data & 0 \\
        Software project & Software design & 1 \\
        Image quality & Image processing & 1 \\
        Autonomous vehicle & Autonomic computing & 1 \\
        Email & Letter & 2 \\
        Distributed computing & Distributed system & 2 \\
        Video processing & Digital image processing & 2 \\
        Computational biology & Computational science & 3 \\
        Static site generator & Web framework & 3 \\
        Operating system & Computer science & 3 \\
        Graphical user interface & User interface & $>$3 \\
        Linear regression & Regression analysis & $>$3 \\
        Unsupervised learning & Machine learning & $>$3 \\
        \bottomrule
        \end{tabular}
\end{table}

\begin{boxH}
Human evaluators agreed on the correctness of 65\% of the taxonomy's \hypohyper pairs, with moderate inter-annotator agreement. This indicates that the final taxonomy is largely accurate, but it contains some ambiguous relationships.
\end{boxH}

\subsection{LLM Evaluation}
GPT evaluated all 929 \hypohyper pairs, assigning a score of 0 to 54\% of the pairs, significantly higher than the 30\% assigned by humans. This difference suggests that GPT is more conservative in assessing pair correctness.

In terms of agreement, including GPT in the analysis resulted in a drop in Krippendorff's Alpha from 0.46 (human-only) to 0.38, indicating that GPT's evaluations diverged more from those of the human annotators. This lower agreement highlights the challenges of incorporating automated models in tasks that require nuanced judgment. Nonetheless, GPT’s performance provides valuable insights, complementing human evaluation by offering a broader perspective, albeit with higher rejection rates.

\begin{boxH}
GPT, being more conservative than human annotators, rejected more pairs, highlighting differences in evaluation criteria between LLMs and human reviewers.
\end{boxH}

\section{Discussion}
\label{sec:discussion}

This section discusses key aspects of the taxonomy construction and evaluation process. 


\subsection{Metrics}
\label{sec:_disc_metrics}
All metrics aid in optimizing the taxonomy, but some were prioritized during the evaluation phase due to their significant impact on usability and relevance. Key metrics include:

\begin{itemize}
    \item \textit{\# New Nodes} reflects the taxonomy’s ability to expand and adapt, ensuring it remains comprehensive and mirrors domain growth. 
    \item \textit{Density} ensures that the nodes within the taxonomy are well-connected. This facilitates user navigation and enhances the overall user experience. 
    \item \textit{Diameter} affects the navigability of the taxonomy. A smaller diameter indicates that users can access related concepts more quickly and efficiently. 
    \item \textit{\# Unlinked} represents gaps in the taxonomy's completeness. A higher number of unlinked nodes indicates that important terms will be missing, reducing the taxonomy’s coverage and relevance to users' needs.
\end{itemize}


In contrast, the metrics \textit{\# Roots}, \textit{\# Components}, and \textit{\# Cycles} are less critical, as they do not directly impact the taxonomy's usability or relevance to the same extent. \textit{\# Roots} refers to the number of primary categories or top-level nodes in the taxonomy. While it is important to maintain a balanced and logical number of root nodes, the overall functionality of the taxonomy is less dependent on this metric as long as the structure remains coherent and easy to navigate. \textit{\# Components} represents the number of disconnected sub-taxonomies or independent sections within the overall taxonomy. Although minimizing the number of components can contribute to a more unified structure, the presence of multiple components is not necessarily detrimental if they are well-organized. Finally, \textit{\# Cycles} can be easily removed with post-processing.

\subsection{Datasources}
\label{sec:_disc_datasources}
We initially hypothesized that only one datasource would be needed and that we would choose the best one out of the three. However, the results show that each has unique characteristics, and combining them yields the best performance.

The CSO model produces taxonomies with minimal variance across most metrics, a low number of new and total terms, and a high count of unlinked terms, reflecting a dense, well-connected structure defined by a single component. In practical applications, a CSO model-generated taxonomy would feature a robust network of interconnected terms, each with explicit semantic relationships. However, The resulting taxonomy has limited usability due to data constraints and missing key terms needed for comprehensive knowledge representation.

Moving to Wikidata, generated taxonomies have a larger number of nodes and a more complex structure characterized by many sparse sub-taxonomies. In real-world applications, these characteristics suggest that Wikidata-based taxonomies could be particularly useful in domains requiring extensive coverage of terms and relationships. However, their complexity and sparsity require additional refinement to ensure efficient querying and usability by linking the sub-taxonomies together and explicitly expressing all relationships in the graph.

The LLM-generated taxonomies provide good term coverage and maintain a lean inclusion of terms; they suffer from structural weaknesses, such as numerous roots, a lack of depth, and a more global structure. In practical implication, this results in a worse version of Wikidata, with many sparse and shallow clusters that require more refinement.

These different characteristics result in each datasource contributing to the final taxonomy in varying amounts. For example, in the final taxonomy, CSO contributes 256 terms and 385 edges, Wikidata provides 398 terms and 497 edges, while LLM contributes 67 terms and 47 edges. Note that terms can belong to more than one datasource if they have edges created from multiple sources.




\subsection{LLMs as Evaluator}
\label{sec:_disc_llms}
While the agreement between human annotators and LLMs is lower than just among the human annotators, LLMs can be beneficial as they lower costs. Therefore, it can automatically assess the quality of the generated taxonomies concerning the semantics and not just the structure. In the case of an LLM-generated taxonomy, a different LLM could be used in a consensus~\cite{chen2023multi} setup, discussion~\cite{chan2024chateval, want2024discussion} setup, or simple evaluation. However, it is still important to perform a manual evaluation once the final taxonomy is created.

\subsection{Construction Process}
\label{sec:_disc_process}

Regarding the process, these aspects should be considered:
\begin{itemize} 
    \item \textbf{Lack of seed terms:} When no seed terms are available, one option is to consult prior work, which is the approach we adopted. However, based on our findings, we also recommend performing seed cleaning to eliminate irrelevant terms and ensure a more focused output. 
       
    \item \textbf{Lack of domain-specific ontology:} If a domain-specific data source is unavailable, or only a related source can be found (as in our case), the proposed method remains effective. However, relying on more general resources like KB or LLMs can introduce additional noise, such as generating excessive new nodes or taxonomies with a less cohesive structure.
    
    \item \textbf{Datasource quality:} While the proposed approach is largely automated, it remains essential to assess the quality of the data source manually before using them rather than relying solely on quantitative metrics.

    \item \textbf{Evaluation:} Human evaluation is time-consuming and not always feasible, so alternatives like LLMs can be used, though with lower precision.
\end{itemize}



\section{Threats to Validity}
\label{sec:threats}

We present the \emph{construct validity}, \emph{internal validity}, \emph{external validity}, and \emph{reliability} that we encountered during our study, and we discuss how we mitigated them.

\paragraph{Construct Validity}
Various threats can affect the construct validity of the proposed approach. 
First, term selection principles may introduce bias if not clearly defined. To address this, we prioritise relevance to the subject area, user interest, and the structural integrity of the taxonomy, ensuring the inclusion of key terms and proper relationships.

Furthermore, when aligning GitRanking terms with those from datasources, incorrect terms or relationships may be included. This is mitigated by applying a semantic check for CSO to filter irrelevant terms. For Wikidata, QIDs prevent this issue, though for LLMs, the challenge remains unaddressed.

Finally, seed terms from GitRanking are manually vetted, reducing subjectivity in selection.

\paragraph{Internal Validity}
Internal validity threats arise from biases in hyperparameter selection, data source variations, and potential errors in cleaning and post-processing. Hyperparameter choices, like term relevance thresholds, may affect taxonomy accuracy, while removing cycles or irrelevant terms could impact quality. These threats are mitigated by using TOPSIS for configuration selection, ensuring robustness and reproducibility through multiple evaluation metrics.

\paragraph{External Validity}
Focusing on a single datasource could limit the methodology's applicability to other domains. To mitigate this, the approach uses multiple datasources, from domain-specific ontologies to general knowledge bases, enhancing coverage and reducing reliance on one source. The method is adaptable by switching ontologies across fields, and the ensemble method, combining multiple taxonomies, creates a more stable structure, reducing bias from individual sources.

\paragraph{Reliability}
The taxonomy construction process may not be reproducible due to the use of LLMs, which can generate different results on repeated trials, and the close-source ones can change without notice. We mitigate this by using versioned models and providing prompts.

\section{Related Work}
\label{sec:sota}


This section covers background on two relevant areas: software categorization and taxonomy construction.

\subsection{Software Categorisations}
There have been various proposed categorisations through the years; we present a summary in this section.

The first categorization has been proposed alongside MUDA~\cite{Kawaguchi2006MUDABlue}, a Latent Semantic Analysis (LSA) based approach for categorizing a project into 6 categories. In ClassifyHub~\cite{soll2017classifyhub}, the authors used the InformatiCup 2017\footnote{\href{https://github.com/informatiCup/informatiCup2017}{https://github.com/informatiCup/informatiCup2017}} dataset, which uses seven categories. Sharma et al.~\cite{sharma2017cataloging} proposed a dataset with 22 labels. 

Moving to approaches that use a subset of other categorizations, LACT~\cite{tian2009lact} uses 6 categories that they selected from SourceForge. Another taxonomy based on SourceForge categories has been proposed by Vasquez \etal~\cite{vasquez2014api}, which uses 22 categories; moreover, the examples are annotated with multiple labels. Moving towards bigger categorizations, LeClair et al.~\cite{leclair2018neural} worked on a dataset of C/C++ projects from the Debian package repository with 75 categories.  

More recent approaches have moved to bottom-up categorization from GitHub~\cite{izadi201repologue,sipio2020naive}. The labels are picked from a subset of the ones developers use to annotate their projects. Each developer can define these without a pre-defined list. Repologue~\cite{izadi201repologue} uses a subset of 228, while Di Sipio \etal~\cite{sipio2020naive} uses a smaller subset of 134.

\subsection{Taxonomy Construction}
Taxonomy construction is an essential task across various domains. It serves as the backbone for applications such as web search, e-commerce, and knowledge representation. This task involves organizing concepts into hierarchical structures that reflect their semantic relationships. 

WordNet~\cite{miller1995wordnet}, an early knowledge base, was carefully curated by experts to map out semantic relationships between concepts. Despite offering structured, high-quality data, this manual method has significant drawbacks: it's time-consuming, resource-heavy, and unsuitable for processing larger data. Consequently, there is a growing demand for automated solutions to build taxonomies at scale.

Traditionally, many taxonomy construction methods have relied on pattern-based techniques to identify hypernyms within documents~\cite{hearst1992automatic, shang-etal-2020-taxonomy, chen-etal-2021-constructing}. These approaches check whether, given a pair of terms, they appear in the same sentence and satisfy a particular pattern.

Other popular approaches are clustering-based, such as TaxoGen~\cite{zhang2018taxogen}, NetTaxo~\cite{shang2020nettaxo}, and CoRel~\cite{huang2020corel}. These approaches typically involve an iterative process where semantic vectors (\ie embeddings) for terms of interest are learned. The workflow involves repeatedly clustering, refining embeddings, and re-clustering to create increasingly discriminative representations. While effective, this process is data-intensive and computationally demanding, especially when new terms are introduced, necessitating a complete rerun of the algorithms. Other approaches, like~\cite{song2015taxonomy}, use a mix of bayesian approaches and clustering. More recently, in \cite{MENG2024112405}, authors proposed a novel contrastive learning approach to refine relationships in subcomponents identified via KNN clustering.

The shift towards using LLMs approaches also reached taxonomy construction. Various approaches have been proposed\cite{wan2024tnt_llm, chen2023llm_taxonomy, gunn2024taxonomy_llm,zeng2024col}. They leverage LLM's ability to understand and generate natural language to infer semantic relationships even in zero-shot scenarios. \cite{gunn2024taxonomy_llm} use LLMs to generate a taxonomy from an initial set of high-level terms in a top-down approach. Similarly, in~\cite{zeng2024col}, the authors use an iterative approach to construct a taxonomy from a small set of examples given a list of terms and a root in a top-down fashion. In~\cite{chen2023llm_taxonomy}, they compare fine-tuning LLMs and prompting for the taxonomy construction task. Lastly, \cite{wan2024tnt_llm} use LLMs to extract \hypohyper pairs from large amounts of textual data.

\section{Conclusions and Future Works}
\label{sec:conclusions}
In this work, we proposed an approach for automatically creating domain-specific taxonomies by leveraging preexisting resources. Unlike prior methods, which often rely on large amounts of textual data or exclusively on large language models, our approach combines preexisting ontologies, knowledge bases, and LLMs. We evaluate this approach in the context of the software application domain.

Quantitative results indicate that each datasource offers unique strengths and weaknesses, with inherent trade-offs regarding term addition and edge creation. Therefore, we harvest these distinct strengths of each source via an ensemble approach to reduce noise and obtain higher-quality results.

Results from human evaluation show that 65\% of the evaluated term pairs were considered correct by at least two annotators, with a moderate level of agreement (0.46). In addition to human evaluations, we also assessed the taxonomy using LLMs, allowing us to measure consistency and accuracy from both automated and human perspectives. This dual evaluation highlights the complementary role of human expertise and machine learning in refining taxonomy creation.

Future work should focus on refining the ensemble techniques
and explore additional datasources to  enhance the taxonomy’s accuracy and completeness. Furthermore, we plan to integrate our taxonomy into the AutoFL tool~\cite{sas2024autofl}, enabling the creation of a dataset annotated by the taxonomy and expanding on the dataset from~\cite{sas2023weak}. This will help evaluate the taxonomy's benefits for software comprehension and serve as a benchmark for improving representation learning approaches, such as embedding models.

We will also explore the taxonomy’s application for code summarization, generation, and search. Additionally, we aim to research its impact on developing more effective software engineering tools and conduct user studies to assess its usability and effectiveness in real-world scenarios.



\section*{Conflict of Interest}
The authors declare that they have no conflict of interest.

{\scriptsize
\bibliography{references}}

\begin{IEEEbiography}[{\includegraphics[width=1in,height=1.25in,clip,keepaspectratio]{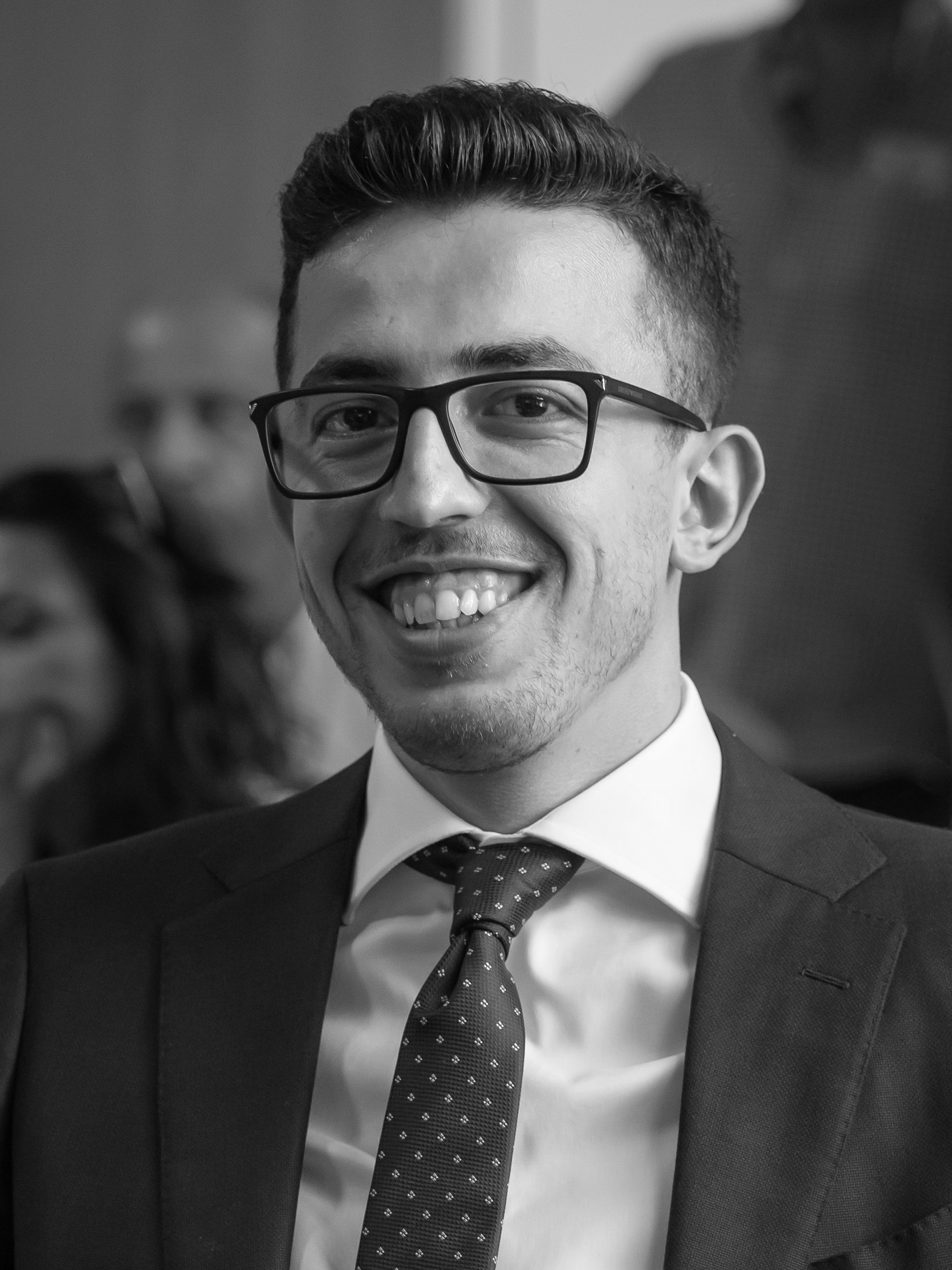}}]{Cezar Sas}
received the M.Sc. degree in Computer Science from the University of Milano-Bicocca, Milan, Italy in 2019. He is currently pursuing a Ph.D. in Computer Science at the University of Groningen, Groningen, The Netherlands. His research interests focus on developing methods for extracting actionable insights from unstructured data. Currently, he is focused on improving software classification.
\end{IEEEbiography}

\begin{IEEEbiography}[{\includegraphics[width=1in,height=1.25in,clip,keepaspectratio]{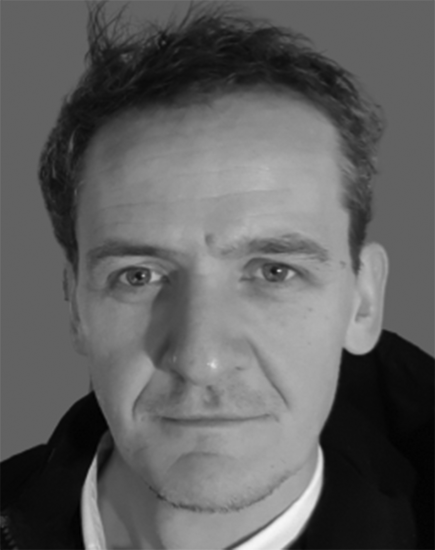}}]{Andrea Capiluppi} received the M.Sc. degree in management engineering and the Ph.D. degree in software engineering from Politecnico di Torino, Torino, Italy, in 2001 and 2005, respectively.
He is currently an Associate Professor with the University of Groningen, Groningen, The Netherlands, where he has established an ecosystem of Dutch industrial partners, within which he develops software solutions as part of the software engineering course that he manages. His research focuses on empirical software engineering..
\end{IEEEbiography}

\vfill




\newpage
\appendix
\subsection{Prompts}
\label{apx:prompts}

Prompt used for the construction of the taxonomy following~\cite{gunn2024taxonomy_llm}:

\begin{llmprompt}[label=llm_2]
\justifying
\texttt{
You are a helpful assistant tasked to pair terms to the hypernym to which they should belong. If it does not belong to any, answer None. \\
Given a term, provide the hypernym for the term. The hypernym should be a term from the taxonomy. \\
Multiple answers are allowed, and should be separated by a comma. Keep the answer concise, in CSV format, without any extra. \\
For example: \\
parent1\\
parent1,parent2,parent3 \\
None\\
This is the list of possible terms: \\
\{taxonomy\}\\
What is the hypernym of \{term\}?
}
\label{prompt:llm2}
\end{llmprompt}

Prompt used for the evaluation of the taxonomy following~\cite{gao2023rate}:

\begin{llmprompt}[label=llm_3]
\justifying
\texttt{
You are an expert in domain relationships and knowledge categorization. Your task is to analyze pairs of terms and determine their relationship based on the following criteria:\\
    - 1 (Subdomain Relationship): One term is a specific subdomain or subset of the other.\\
    - 0 (No Relationship): The terms have no significant relationship. \\
    For each pair of terms provided, identify and categorize their relationship. Only provide the classification (0, or 1) without any explanation. \\
    Are the terms in the pair related as subdomain or unrelated?\\
    \{pair\}
}
\end{llmprompt}

\end{document}